\documentclass[apj,12pt,useAMS]{emulateapj}

\slugcomment{}

\shorttitle{On the electron energy distribution index of \emph{Swift} GRBs}
\shortauthors{Curran et al.}

\begin{document}

\title{On the electron energy distribution index of \\ \emph{Swift} Gamma-Ray Burst afterglows}


\author{P.A. Curran\altaffilmark{1}\altaffilmark{*}, 
P.A.~Evans\altaffilmark{2}, 
M.~de~Pasquale\altaffilmark{1},
M.J.~Page\altaffilmark{1},
A.J.~van~der~Horst\altaffilmark{3} }

\altaffiltext{*}{{\tt pac@mssl.ucl.ac.uk} }
\altaffiltext{1}{Mullard Space Science Laboratory, University College of London, Holmbury St Mary, Dorking, Surrey RH5\,6NT, UK}
\altaffiltext{2}{Department~of~Physics~and~Astronomy, University~of~Leicester, University~Road, Leicester~LE1~7RH, UK }
\altaffiltext{3}{NASA Postdoctoral Program Fellow, NSSTC, 320 Sparkman Drive, Huntsville, AL 35805, USA }

\begin{abstract}
The electron energy distribution index, $p$, is a fundamental parameter of the synchrotron emission from a range of astronomical sources. 
Here we examine one such source of synchrotron emission, Gamma-Ray Burst afterglows observed by the \emph{Swift} satellite. Within the framework of the blast wave model, we examine the constraints placed on the distribution of $p$ by the observed X-ray spectral indices and parametrise the distribution. 
We find that the observed distribution of spectral indices are inconsistent with an underlying distribution of $p$ composed of a single discrete value but consistent with a Gaussian distribution centred at $p = 2.36$ and having a width of $0.59$. Furthermore, accepting that the underlying distribution is a Gaussian, we find the majority ($\gtrsim 94$ percent) of GRB  afterglows in our sample have cooling break frequencies less than the X-ray frequency.
\end{abstract}



\section{Introduction}\label{section:introduction}

The afterglow emission of Gamma-Ray Bursts (GRBs) is generally described by the blast wave model  \citep{rees1992:mnras258,meszaros1998:apj499} which details the temporal and spectral behaviour of the emission that is created by external shocks when a collimated ultra-relativistic jet ploughs into the circumburst medium, driving a blast wave ahead of it. 
Fundamental to this model is the electron energy distribution index, $p$; a characteristic parameter of the process by which the electrons are accelerated to relativistic speeds and by which they radiate via synchrotron emission. This acceleration mechanism, common to many astronomical jet sources (as well as particle acceleration in the solar wind and supernovae, and the acceleration of cosmic rays), is thought to be Fermi diffusive shock acceleration \citep{fermi1954:ApJ119} due to the passage of an external shock  \citep{blandford1978:ApJ221,rieger2007:Ap&SS309} after which the energy of the electrons, $E$, follows a power-law distribution, $\mathrm{d}N \propto E^{-p} \mathrm{d}E$, with a cut-off at low energies. 
This is consistent with recent PIC simulations \citep{spitkovsky2008:ApJ682} and Monte Carlo models \citep{achterberg2001:MNRAS328,ellison2002:APh18,lemoine2003:ApJ589} but at odds with others \citep{niemiec2006:ApJ641,niemiec2006:ApJ650}.

The blast wave model describes how synchrotron emission from relativistic electrons produces a smoothly-broken, broad band spectrum that is well characterised by a peak flux and three, time evolving, break frequencies (peak frequency, $\nu_{\mathrm{m}}$; cooling frequency, $\nu_{\mathrm{c}}$; synchrotron self-absorption frequency, $\nu_{\mathrm{a}}$) as well as the electron energy distribution index, $p$ \citep{sari1998:ApJ497,granot2002:ApJ568}. The spectrum is divided into four regimes by the three break frequencies and within each regime the spectrum is asymptotically described by $F_{\nu} \propto \nu^{-\beta}$, where the spectral index, $\beta$, is a function of $p$ only. 
By comparing the observed X-ray spectra to the predicted asymptotic values of the synchrotron spectra, we can extract information about the electron energy distribution index, $p$, which is dependent only on the underlying micro-physics of the acceleration process.
Some (semi-)analytical calculations and simulations indicate that there is a nearly universal value of $\sim 2.2-2.4$ (e.g., \citealt{kirk2000:ApJ542,achterberg2001:MNRAS328,spitkovsky2008:ApJ682}) though other studies suggest that there is a large range of possible values for $p$ of $1.5-4$ \citep{baring2004:NuPhS136}.

Observationally, different methods have been applied to samples of BATSE, \emph{BeppoSAX} and \emph{Swift} bursts which reached the conclusion that the observed range of $p$ values is not consistent with a single central value of $p$ 
\citep{chevalier2000:ApJ536,panaitescu2002:ApJ571,shen2006:MNRAS371,starling2008:ApJ672,curran2009:MNRAS395}. The latter three showed that the width of the parent distribution is $\sigma_{p} \sim 0.3-0.5$. 
However, in the studies so far there have been some limitations: multi-wavelength studies \citep{chevalier2000:ApJ536,panaitescu2002:ApJ571,starling2008:ApJ672,curran2009:MNRAS395} suffer from limited samples ($\lesssim 10$ sources each) with sufficient temporal and spectral observations, while those studies that rely on X-ray afterglows alone are subject to a large uncertainty because the position of the cooling frequency, relative to the X-ray, is unknown. The only X-ray study of \emph{Swift} afterglows so far \citep{shen2006:MNRAS371} used a very limited sample of spectral indices ($\sim 30$), dictated by the number of GRBs observed by \emph{Swift} at the time, to estimate the distribution of $p$. Neither, they did not take a statistical approach to the position of the cooling frequency relative to the X-ray regime, as we do here.
We interpret a much larger ($\sim 300$) and, statistically, more significant sample of \emph{Swift} observed GRB afterglows, to constrain the {\it distribution} of the values of electron energy distribution index, $p$.
In \S\ref{section:method} we introduce our method while in \S\ref{section:results} we present the results of our Monte Carlo analyses and their implications in the overall context of GRB observations and particle acceleration in general. 
We summarise our findings in \S\ref{section:conclusion}.
All uncertainties are quoted at the $1\sigma$ confidence level.


\section{Method}\label{section:method}

\begin{figure}
\epsscale{.80}
\includegraphics[scale=.35,angle=-90]{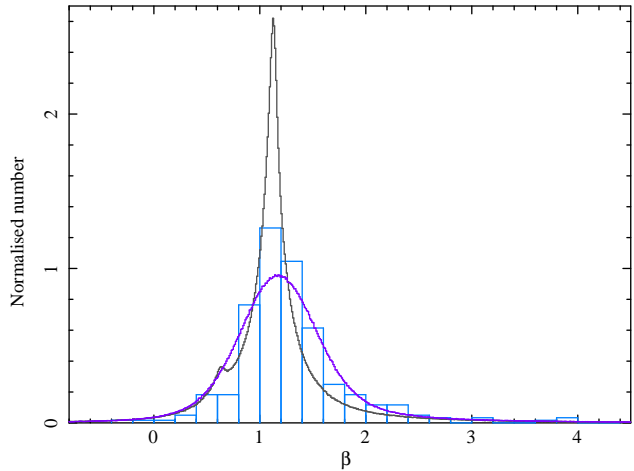} 
\caption{Normalised histogram of the data  (301 measurements) overlaid with high-resolution normalised histograms of the synthesized data sets ($10^4 \times 301$ data points) from the most likely parameters of a single discrete $p$ (grey line) and a Gaussian distribution of $p$ (blue line) as detailed in Table\,\ref{table:results}.}
  \label{fig:hist} 
\end{figure}

Our general method is to constrain the electron energy distribution index, $p$, from the values of the X-ray spectral indices, $\beta_{{\rm X}}$ or $\beta$,
observed by the \emph{Swift} XRT \citep{burrows2005:SSRv120} and detailed by \citet[Table\,7]{evans2009:MNRAS397}.
A normalised histogram of the spectral indices of these 301 GRB spectra is plotted in figure \ref{fig:hist}. 
We derive $p$ from the spectral index as opposed to the temporal index because for a given spectral index, assuming the asymptotic limit, there are only two possible values of $p$ depending on whether the cooling frequency, $\nu_{{\rm c}}$ is less than or greater than the X-ray regime, $\nu_{{\rm X}}$, while for a given temporal index there are multiple possible values which are model dependent (e.g., the simple  blast wave model \citep{rees1992:mnras258,meszaros1998:apj499}, or modifications thereof (e.g., \citealt{granot2006:MNRAS366,genet2007:MNRAS381,ghisellini2007:ApJ658,uhm2007:ApJ665L}); for a discussion on the choice of X-ray spectral index to derive $p$ see \citealt{curran2009:MNRAS395}).
In accordance with synchrotron emission predicted by the blast wave model, we ascribe the behaviour of the unabsorbed X-ray spectrum to be a single power law where the flux goes as: $F_{\nu}(\nu) \propto  \nu^{-\beta}$ and $\beta$ is the spectral index.
Under the standard assumptions of slow cooling and adiabatic expansion of the blast wave, the electron energy distribution index is related, in the asymptotic limit, to the spectral index by either $p = 2\beta$ ($\nu_{\mathrm{c}} < \nu_{\mathrm{X}}$) or $p = 2\beta +1$ ($\nu_{\mathrm{c}} > \nu_{\mathrm{X}}$) (e.g., \citealt{granot2002:ApJ568});  implying a difference between the slopes of the two spectral regimes of $\Delta\beta = 0.5$.  
Throughout we use the regime probability, $X$, as the probability that the cooling frequency is less than the frequency of the X-ray regime (i.e., $\nu_{\mathrm{c}} < \nu_{\mathrm{X}}$) and $1-X$ is the probability that $\nu_{\mathrm{c}} > \nu_{\mathrm{X}}$. We neglect the case where the cooling frequency may be passing through the X-ray regime (since there is no sign of spectral evolution in the sample) and the cases where the peak frequency, $\nu_{\mathrm{m}}$, or self absorption frequency, $\nu_{\mathrm{a}}$, is greater than the X-ray regime as this is not observed in late time afterglows.

To parameterise the underlying distribution of the electron energy distribution index, $p$, from the X-ray spectral indices observed by \emph{Swift} we use a maximum likelihood Monte Carlo method. This method uses a maximum likelihood fit to return the \emph{most likely} parameters of the assumed underlying model, the errors on which are estimated via a Monte Carlo 
error analysis. Another Monte Carlo analysis tests the probability that the observed distribution of spectral indices could be obtained from an underlying distribution of $p$ described by the most likely parameters. 

In this method we first assume a model for the underlying distribution of $p$  which we transform into a distribution in spectral index space, via the regime probability, $X$. We convolve this distribution with the measured probability of the data set and calculate the log-likelihood of the parameters of the underlying model (see appendix\,\ref{section:prob.functions}). To estimate the most likely parameters of the underlying model and the regime probability, $X$, we minimise the log-likelihood (equates to maximising the likelihood) using the \emph{simulated annealing} method ($\S$ 10.9 of \citealt{press1992:nrca}, and references therein). Uncertainties of the fit parameters are estimated via a Monte Carlo analysis, whereby the observed data are randomly perturbed within their (asymmetric) Gaussian errors and refit multiple times ($10^4$); the standard deviation of returned most likely parameters is used as a measure of the uncertainties. 
We then find the probability that the observed distribution of spectral indices could have originated from an underlying distribution of $p$ based on the most likely parameters estimated from the log-likelihood minimisation. We do so by generating $10^4$ synthetic data sets drawn randomly from the underlying probability distribution of $p$, randomly transformed into spectral index space via the regime probability, $X$, and further randomly perturbed within the (asymmetric) Gaussian errors of each of our original data points. 
Each of these synthetic data sets is fit as above and the value of the log-likelihoods recorded; one would expect that  if the data are consistent with the underlying model, at the $1\sigma$ level, the original log-likelihood value should fall within the $1\sigma$ distribution of the synthetic values.
With $10^4$ synthetic data sets we can measure the percentage to an accuracy of two decimal places and rule out chance agreements to the $4 \sigma$ level; the $\sim 2 \times 10^5$ synthetic data sets required to rule out chance agreements at the $5 \sigma$ level was considered too costly, computationally.

There are two underlying models, or hypotheses, regarding the data that we want to test: that the observed distribution of spectral indices, $\beta$, can be obtained from an underlying distribution of $p$ composed of $i)$ a single discrete value (SD$p$) and $ii)$ a Gaussian distribution (GD$p$).  Details of these models and their log-likelihoods are discussed in the Appendix, \ref{section:prob.functions}.


\section{Results and Discussion}\label{section:results}

\begin{table*}
\begin{center}
\caption{Results of our likelihood fits of the observed spectral index distributions to a single discrete value (SD$p$) and a Gaussian distribution (GD$p$) of $p$.  }    \label{table:results} 

    \begin{tabular}{l l l l l l } 
\tableline\tableline
     model  & $p$      &  $\sigma_{p}$ & $X$ &  $l$ & \% \\ 
\tableline
      SD$p$     & 2.25   &   --    &  0.93 & 1115  & \\ 
                & (2.24$\pm$0.03)  &   --    &  (0.91$\pm$0.02) & (1287$\pm$77) &  \\
                &  [2.253$\pm$0.011]     &    --        &  [0.93$\pm$0.02] &  [58$\pm$28] & 100.00 \\ 
         & &  &  &  &  \\
      GD$p$  &  2.36 &  0.590   &  1.000  & 347  & \\ 
            &  (2.40$\pm$0.03)  &  (0.600$\pm$0.007)  &  (0.991$\pm$0.013)  & (478$\pm$25)  & \\ 
         &      [2.36$\pm$0.05]  &  [0.590$\pm$0.012]  &  [0.99$\pm$0.02]  &   [382$\pm$24] & 7.01 \\
\tableline

\end{tabular}
\tablecomments{The most likely values of electron energy distribution index, $p$, the related Gaussian scatter, $\sigma_{p}$, the probability that the cooling frequency is less than the X-ray frequency ($\nu_{\mathrm{c}} < \nu_{\mathrm{X}}$), $X$, and the log-likelihood of that fit, $l$.  Values in brackets are the average and error values from the Monte Carlo error analysis of perturbed data sets, while those in square brackets are the  average and standard deviations returned from the Monte Carlo analysis of synthesized data sets.  \% is the percentage of synthesized data sets with a better fit than the original. }
\end{center}
\end{table*}

\subsection{Distribution of $p$}\label{p-distribution}

The results of our analysis are detailed in Table\,\ref{table:results} which shows our most likely parameters for electron energy distribution index, $p$, the related Gaussian scatter, $\sigma_{p}$, the probability that the cooling frequency is less than the X-ray frequency ($\nu_{\mathrm{c}} < \nu_{\mathrm{X}}$), $X$, and the log-likelihood of that fit, $l$.  Values in brackets are the average and error values from the Monte Carlo error analysis of perturbed data sets, while those in square brackets are the  average and standard deviations returned from the Monte Carlo analysis of synthesized data sets.  \% is the percentage of synthesized data sets with a better fit than the original.

\begin{figure}
\epsscale{.80}
\includegraphics[scale=.35,angle=-90]{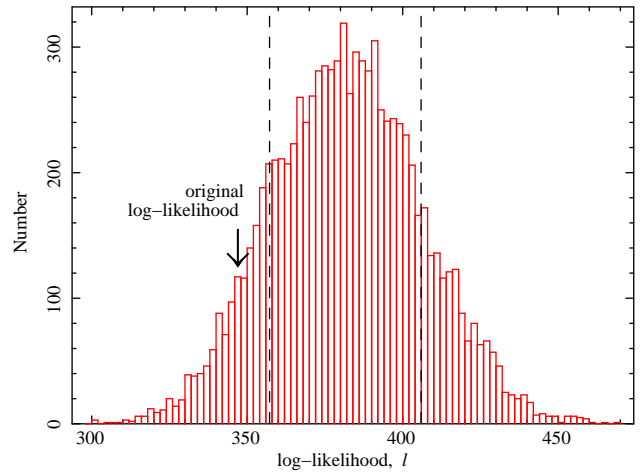} 
\caption{Histogram of the log-likelihood values of the $10^4$ synthesized data sets generated from the most likely parameters, assuming a Gaussian distribution of $p$, as detailed in Table\,\ref{table:results}. The log-likelihood of the original fit to the unperturbed data is marked along with the $1\sigma$ distribution (dashed lines), showing that the data are consistent with the model at the $1.5\sigma$ level.}
  \label{fig:l-dist} 
\end{figure}

Though our most likely discrete value of $p = 2.25$  agrees well with the predicted universal value of $p \sim 2.2-2.3$ (e.g., \citealt{kirk2000:ApJ542,achterberg2001:MNRAS328}), it is comprehensively rejected by our tests; 
the hypothesis that the observed distribution of spectral indices, $\beta$, can be obtained from an underlying distribution of $p$ composed of a single discrete value (SD$p$) is rejected at the $4\sigma$ level as all synthesized data sets had better log-likelihoods. 
The hypothesis that the observed distribution can be obtained from an underlying Gaussian distribution (GD$p$) centred at $p = 2.36$, having a width of $0.590$ and regime probability, $X = 1.000$,  is acceptable at the $1.5 \sigma$ level (Figure\,\ref{fig:l-dist}).  
As a visual aid, we compare the normalised histogram of the observed data with the high-resolution normalised, average histograms of the $10^{4}$ synthesized data sets (Figure\,\ref{fig:hist}). Note that the SD$p$ model is clearly a poor fit and exhibits a secondary peak at $\beta \sim 0.6$ due to the fact that $X = 0.96$  for the likelihood fit of that model to the observed spectral indices. 
This result confirms the results from previous small-sample GRB afterglow studies \citep{shen2006:MNRAS371,starling2008:ApJ672,curran2009:MNRAS395} as regards the non-universality of $p$, the central value at $p \sim 2.0 - 2.5$, and the width of the distribution of $\sigma_{p}\sim 0.3 - 0.5$.  However, our results are based on a sample of bursts an order of magnitude larger than these studies and  we took a statistical approach to the position of cooling frequency relative to the X-ray, by using the regime probability, $X$.

Given that the value of $p$ is not universal, it may be possible that it changes suddenly or evolves gradually with time or radius even in a single event as environmental shock parameters (e.g., magnetic field, ambient density) change or evolve (e.g., \citealt{hamilton1992:ApJ398,vlahos2004:ApJ608,kaiser2005:MNRAS360}). It is also possible that different components of a structured jet \citep{meszaros1998:apj499,kumar2003:ApJ591}, multi-component jet \citep{burger2003:Natur426,huang2004:ApJ605} or jet-cocoon \citep{zhang2003:ApJ586} could have different values of $p$. If either are the case our derived values of $p$ should be considered time averaged values of the parameter, though a change or evolution of $p$ should be observable as a change or evolution of the synchrotron spectral index, $\beta$, and no significant example of such an evolution has been observed in GRB afterglows.

\subsection{Limits on regime probability, $X$}\label{limits}

\begin{figure}
\epsscale{.80}
\includegraphics[scale=.35,angle=-90]{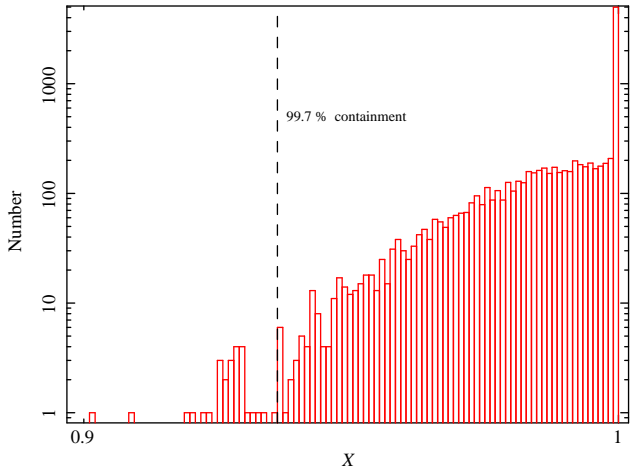} 
\caption{Histogram (on a log-log plot) of the values of $X$ returned by the Monte Carlo error analysis, assuming a Gaussian distribution of $p$. The dashed vertical line shows the 99.7\% containment at $X=0.935$ which we take as a measure of the lower limit on $X$.  }
  \label{fig:X_limits} 
\end{figure}

Though previous multi-band (optical -- X-ray) studies (e.g., \citealt{panaitescu2002:ApJ571,starling2008:ApJ672,curran2009:MNRAS395}) have shown that the cooling break frequencies of a number of GRB  afterglows are greater than the X-ray frequency ($\nu_{\mathrm{c}} > \nu_{\mathrm{X}}$), here, accepting that the underlying distribution of $p$ is a Gaussian, we find that this number is consistent with zero ($X = 1$).  From our Monte Carlo error analysis (where we refit the randomly perturbed data multiple times) we can plot the distribution of possible values of $X$ (Figure\,\ref{fig:X_limits}; note that this is a log-log plot). The distribution of the errors is clearly not Gaussian so we should not use the standard deviation as a measure of error as we have in Table\,\ref{table:results}. We can however place a  nominal $3\sigma$ lower limit on the value of $X$ by estimating the value at which there is 99.7 percent containment; we find that this is at  $0.935 < X < 0.936$, compared to the nominal $1\sigma$ (68.2 percent containment) at $0.988 < X < 0.989$. 
Any value of X less than this limit is inconsistent with the distribution of the data and 
would produce a clear secondary peak at $\beta = (p-1)/2$, not observed in the data  (Figure\,\ref{fig:hist}). 
To avoid this peak, an extremely wide distribution is needed, much wider than the spread of the observed data.

Hence, the upper limit on the percentage of GRBs in our sample where 
the cooling frequency is greater than the X-ray frequency ($\nu_{\mathrm{c}} > \nu_{\mathrm{X}}$) is approximately 6.5 percent, or $\sim20$ out of 301 GRBs. This is a discrepancy from the ratio observed in the multi-band studies (e.g., \citeauthor{starling2008:ApJ672,curran2009:MNRAS395}: 5 out of 10 and 2 out of 6, respectively) but not seriously so given the extremely low number statistics of those studies. 
If we create a sample of 6 random bursts from our statistical distribution of 301, 
there would be a non-negligible ($\sim$ 5 percent) chance that 2, or more, 
bursts would have a cooling frequency greater than the  X-ray frequency, 
consistent with the aforementioned \emph{Swift} study (\citeauthor{curran2009:MNRAS395}). 
The study of \citeauthor{starling2008:ApJ672} is based on a sample of \emph{BeppoSAX}, as 
opposed to \emph{Swift}, GRBs  which may have a different limit on the regime probability, $X$. 
An investigation as to why the cooling frequency of afterglows follow this apparent limit, or its implications regarding the distribution of other 
parameters, is beyond the scope of this work.


\section{Conclusion}\label{section:conclusion}

We use the X-ray spectral indices of gamma-ray burst afterglows observed by the XRT aboard \emph{Swift} to parameterise the underlying distribution of the electron energy distribution index, $p$, within the framework of the blast wave model.  The electron energy distribution index is a fundamental parameter of the synchrotron emission from a range of astronomical sources and in this case of the synchrotron emission of GRB afterglows. 
We use a maximum likelihood Monte Carlo analysis to test two hypotheses, namely  that the observed distribution of spectral indices, $\beta$, can be obtained from an underlying distribution of $p$ composed of $i)$ a single discrete value and $ii)$ a Gaussian distribution. We find that the observed distribution of spectral indices are inconsistent with the first hypothesis but consistent with the second, a Gaussian distribution centred at $p = 2.36$ and having a width of $0.59$. Furthermore, if we accept that the underlying distribution is a Gaussian, the majority ($\gtrsim 94$ percent) of GRB afterglows in our sample have a cooling break frequency less than the X-ray frequency.

\acknowledgments

We thank the referee for their comments.  
PAC, PAE, MJP, MdP  acknowledge support from STFC.
AJvdH was supported by an appointment to the NASA Postdoctoral Program at the MSFC, administered by Oak Ridge Associated Universities through a contract with NASA.

{\it Facilities:} \facility{Swift}.


\appendix
\section{Probability functions}\label{section:prob.functions}

The probability of spectral index, $\beta$,  given a measured value, $\beta_i$, with asymmetric errors, $\sigma_{i\pm}$ \citep{starling2008:ApJ672} is 
\begin{eqnarray} \label{eqn:Pi_beta}
\mathrm{P}_i(\beta | \beta_i,\sigma_{i\pm})  & = & 
\frac{1}{\sqrt{2\pi}\, \overline{\sigma_{i\pm}} }
\exp \left[  \frac{-(\beta_i-\beta)^2}{2 \sigma_{i\pm}^{2}}    \right] , 
\end{eqnarray}
where $\overline{\sigma_{i\pm}} =  (\sigma_{i-} + \sigma_{i+} )/2$ and 
$\sigma_{i\pm}= \left\{  
\begin{array}{ll} 
 \sigma_{i-} & (\beta < \beta_i)  \\
 \sigma_{i+} & (\beta \geq \beta_i) 
\end{array} \right. .$ \\

Assume that the distribution of the electron energy distribution index, $p$, can be described by a Gaussian probability:  
\begin{eqnarray} \label{eqn:P_p}
\mathrm{P}(p | \bar{p},\sigma_{p}) & = &
\frac{1}{\sqrt{2\pi}\,\sigma_{p} } \exp \left[ \frac{-(p-\bar{p})^2}{2 \sigma_{p}^{2}}    \right]   , 
\end{eqnarray}
where $\bar{p}$ is the central, or mean, value of the distribution and $\sigma_{p}$ is the standard deviation of the distribution. 
The distribution of the observed X-ray spectral indices, $\beta$, can then be described by a double Gaussian:
\begin{eqnarray} \label{eqn:P_beta_p}
\mathrm{P}(\beta | \bar{p},\sigma_{p}, X) & = &
\frac{1}{\sqrt{2\pi}\,(\sigma_{p}/2) } 
\left(
X \exp  \left[ \frac{-(\beta-\bar{p}/2)^{2}} {2 (\sigma_{p}/2)^{2}} \right] 
+ (1-X)  \exp \left[ \frac{-(\beta-\bar{p}/2+0.5)^{2}} {2 (\sigma_{p}/2)^{2}} \right]  
\right)  , \nonumber\\
\end{eqnarray}
where $X$ is the probability that the the cooling frquency is less than the X-ray frequency ($\nu_{\mathrm{c}} < \nu_{\mathrm{X}}$), and $(1-X)$ is the probability that the cooling frequency is greater than the X-ray frequency ($\nu_{\mathrm{c}} > \nu_{\mathrm{X}}$).  Convolving this probability with the measured probability (equation \ref{eqn:Pi_beta}) gives:
\begin{eqnarray} \label{eqn:Pi_beta_p}
\mathrm{P}_i(\beta | \bar{p},\sigma_{p}, X, \beta_i,\sigma_{i\pm} ) & = &
\frac{1}{\sqrt{2\pi}} \, \frac{1}{\sqrt{(\sigma_{p}/2)^2 +  \overline{\sigma_{i\pm}}^2}} \nonumber\\
 & & \times
\left( X
\exp  \left[ \frac{-(\beta_i-\bar{p}/2)^{2}}{2 ( (\sigma_{p}/2)^2 + \sigma_{i\pm}^2)}  \right] 
+ (1-X)  \exp  \left[ \frac{-(\beta_i-\bar{p}/2+0.5)^{2}}{2 ((\sigma_{p}/2)^{2} + \sigma_{i\pm}^2)} \right]  
\right)  , \nonumber\\
\end{eqnarray} 
where
$\sigma_{i\pm}= \left\{  
\begin{array}{ll} 
 \sigma_{i-} & (\beta_i < \bar{p}/2; ~\mathrm{first\,exponential})  \\
 \sigma_{i+} & (\beta_i \geq \bar{p}/2; ~\mathrm{first\,exponential}) \\
 \sigma_{i-} & (\beta_i < \bar{p}/2-0.5; ~\mathrm{second\,exponential})  \\
 \sigma_{i+} & (\beta_i \geq \bar{p}/2-0.5; ~\mathrm{second\,exponential}) ~ .
\end{array} \right. $ \\
This leads to 3 different permutations  of equation \ref{eqn:Pi_beta_p}, since the combination of $\beta_i < \bar{p}/2-0.5$ and $\beta_i \geq \bar{p}/2$ does not exist.
The log-likelihood of these probability distributions is given by
\begin{eqnarray} \label{eqn:l_p}
l(\bar{p},\sigma_{p}, X) 
& = & -2 \ln L(\bar{p},\sigma_{p}, X) \nonumber\\
& = & -2 \ln \displaystyle\prod_{i=1}^{N} P_i(\beta | \bar{p},\sigma_{p}, X, \beta_i,\sigma_{i\pm}) \nonumber\\
& = & -2  \displaystyle\sum_{i=1}^N   \ln  P_i(\beta | \bar{p},\sigma_{p}, X, \beta_i,\sigma_{i\pm})    \nonumber\\
& = &
N \ln(2\pi) +  
\displaystyle\sum_{i=1}^{N}  
\ln((\sigma_{p}/2)^2 +  \overline{\sigma_{i\pm}}^2)  \nonumber\\
& &  
-2\displaystyle\sum_{i=1}^{N}  
 \ln  \left(
 X \exp  \left[ \frac{-(\beta_i-\bar{p}/2)^{2}}{2 ( (\sigma_{p}/2)^2 + \sigma_{i\pm}^2)}  \right] 
+ (1-X)  \exp  \left[ \frac{-(\beta_i-\bar{p}/2+0.5)^{2}}{2 (  (\sigma_{p}/2)^{2} + \sigma_{i\pm}^2)} \right]  
\right). \nonumber\\
\end{eqnarray}
where $L(\bar{p},\sigma_{p}, X)$ is the likelihood function. 
This is a numerically calculable function that is minimised to find the most likely parameters, errors on which can be estimated via a Monte Carlo error analysis. 
If the distribution of the electron energy distribution index, $p$, can be described by a single discrete value, \\
$\mathrm{P}(p | \bar{p}) = \left\{  
\begin{array}{ll} 
 1 & (p = \bar{p})  \\
 0 & (p \neq \bar{p}) 
\end{array} \right.$,  
then the probability of $\beta$ is 
\begin{equation}
\mathrm{P}(\beta | \bar{p}) = \left\{  
\begin{array}{ll} 
 X & (\beta = \bar{p}/2)  \\
 1-X & (\beta = \bar{p}/2 - 0.5) \\
 0 & (\beta \neq \bar{p}/2 ~\&~ \beta \neq \bar{p}/2 - 0.5) 
\end{array} \right.  
\end{equation}
and the above convolved probabilities and likelihoods hold with $\sigma_{p} = 0$.



\end{document}